\input psfig.sty
\input epsf.sty

\def\mev{\,{\rm Me\kern-0.1em V}}
\def\gev{\,{\rm Ge\kern-0.1em V}}

\documentstyle[12pt]{article}

\renewcommand{\baselinestretch}{1.8}
\begin{document}
\begin{center}
{\Large{\bf Computing Electromagnetic Effects in Fully Unquenched QCD}}\\
\vspace*{.45in}
{\large{A. ~Duncan$^1$, E. ~Eichten$^2$,
and R. ~Sedgewick$^3$}} \\
\vspace*{.15in}
$^1$Dept. of Physics and Astronomy, Univ. of Pittsburgh, 
Pittsburgh, PA 15260\\
$^2$Theory Group, Fermilab, PO Box 500, Batavia, IL 60510\\
$^3$Dept. of Chemistry, Univ. of Pittsburgh, Pittsburgh, PA 15260\\
\end{center}
\vspace*{.3in}
\begin{abstract}
The inclusion of electromagnetic effects in unquenched QCD can be
accomplished using ensembles generated in dynamical simulations
with pure QCD provided the change in the quark determinant induced
by a weak electromagnetic field can be efficiently computed. A
stochastic technique for achieving this in the case of dynamical
domain wall calculations is described.
\end{abstract}
\newpage

\section{Introduction}

 Previous work in quenched QCD \cite{DET1,DET2} has established the
possibility of computing quark masses including electromagnetic effects,
as well as fine structure of hadron multiplets, by explicitly coupling 
the quark fields to a weak electromagnetic U(1) gauge field superimposed
on pure QCD gauge fields generated in a conventional Monte Carlo simulation.
Although the pattern of fine structure revealed in these quenched 
calculations qualitatively matches the experimental splittings for the
meson and baryon octets, a credible quantitative computation clearly 
requires a fully unquenched treatment of both the gluon and photon dynamics.
In this paper we present a technique for including electromagnetic effects
in preexisting ensembles of pure QCD gaugefields generated with dynamical
domain wall quarks. Although we describe the method for domain wall quarks
(preferred in this case for the enhanced chiral symmetry and control over
quark mass renormalization available in this formulation), the technique
applies just as well to other fermion discretizations such as Wilson
quarks (with or without clover improvement).

 To illustrate the basic idea, we note that the path integral for quarks
coupled to both SU(3) gluons (represented by conventional compact link
variables $U$, with action $S_{\rm gl}$)  and an abelian  photon field $A$ 
(with a noncompact action $S_{\rm em}(A) = \frac{1}{4e^2}\sum_{n\mu\nu}(\nabla_{\mu}A_{n\nu}-
\nabla_{\nu}A_{n\mu})^{2}$,\cite{DET1}, 
in Coulomb gauge) can be written
(somewhat schematically):
\begin{eqnarray}
  Z_{SU(3){\rm x}U(1)} &=& \int {\cal D}U{\cal D}A\;\;
{\rm det}(M(U,A))e^{-S_{\rm gl}(U)-S_{\rm em}(A)} \\ 
 &=& \int {\cal D}U\;\; {\rm det}(M(U,A=0))  e^{-S_{\rm gl}(U)}\int {\cal D}A\frac{{\rm det}(M(U,A))}{{\rm det}(M(U,A=0))}e^{-S_{\rm em}(A)}
\end{eqnarray}
where $M(U,A)$ is the matrix defining the lattice quark action in the presence of the
SU(3) field $U$ together with the abelian photon field $A$. Evidently, at least
in principle,  the simulation
of physical observables in the presence of both QCD and electromagnetism can be
effected by superimposing photon fields generated with weight 
$e^{-S_{\rm em}(A)+\ln{\frac{{\rm det}(M(U,A))}{{\rm det}(M(U,0))}}}$ on a pre-existing
ensemble {$U$} generated by a pure SU(3) dynamical QCD simulation. 
This approach will of course only be feasible provided:
\begin{enumerate}
\item The variance of the determinant ratios  
$\ln{(\frac{{\rm det}(M(U,A))}{{\rm det}(M(U,0))})}$
 is tolerably small, and
\item there is a reasonably efficient strategy for computing the above 
determinant ratios for a given gluon field $U$ and a suitable ensemble of
weak photon fields $A$. 
\end{enumerate}
Given the extremely expensive nature of full dynamical QCD calculations,
especially in chirally invariant formulations such as with domain wall
or overlap fermions, it is unrealistic to expect that dynamical configurations
with explicitly simulated photon fields, at a variety of quark masses and
electric charges, will be available in the foreseeable future. Consequently,
a procedure which allows us to exploit dynamical configurations generated
for electrically neutral quarks to study issues of electromagnetic fine
structure would clearly be very useful.

  In Section 2, we show how the determinant ratios appearing in (2) may be computed exactly,
at least for small lattices, for the domain wall formulation of SU(3)xU(1)
gauge theory, using a Lanczos algorithm. Although this approach is not feasible
for large lattices, it gives an exact result on smaller lattices which is
extremely useful in checking the correctness and accuracy of the stochastic
approach to computing determinant ratios described in Section 3. In Section 4
we give further details regarding the computational effort required to extract
determinant ratios at a given level of accuracy. 
  
\section{Exact Computation of Domain Wall Determinants}

  In the domain wall formulation \cite{Sham,Borici} of lattice quark fields,
the matrix $D(U)$ defining the quadratic form for the fermionic action can be
rendered hermitian by premultiplication with the four-dimensional $\gamma_5$
matrix and the time inversion operator $R$ in the fifth dimension (assumed
of extent $L_5$ henceforth), yielding
a matrix $M(U)\equiv R\gamma_{5}D(U)$ which as an $L_{5}$x$L_{5}$  dimensional
block matrix takes the form (for ease of display, taking $L_5$=6)
\[ M = \left(  \begin{array}{cccccc}
   -m_{q}P_{+} & 0 & 0 & 0 & -P_{-} & a_{5}H-\gamma_{5} \\
   0 & 0 & 0 & -P_{-} & a_{5}H-\gamma_5 & P_{+} \\
   0 & 0 & -P_{-} & a_{5}H-\gamma_5 & P_{+} & 0 \\
   0 & -P_{-} & a_{5}H-\gamma_5 & P_{+} & 0 & 0 \\
   -P_{-} & a_{5}H-\gamma_5 & P_{+} & 0 & 0 & 0 \\
   a_{5}H-\gamma_5 & P_{+} & 0 & 0 & 0 & m_{q}P_{-} 
   \end{array} \right)                  \]
where $P_{\pm} \equiv \frac{1}{2}(1 \pm \gamma_{5})$, $m_{q}$ is the bare quark mass,
$a_{5}$ the lattice spacing in the fifth dimension, $H$ the four-dimensional hermitian Wilson operator with an appropriately
chosen domain wall mass (or hopping parameter $\kappa$). 
It is easy to establish the following trace identities
\begin{eqnarray}
  {\rm Tr}(M) &=& 0  \\
  {\rm Tr}(M^2) &=& 12V(L_{5}(1-\frac{a_{5}}{2\kappa})^{2}+L_{5}(1+4a_{5}^{2})+m_{q}^{2}-1)
\end{eqnarray}
where $V$ is the volume of the four-dimensional lattice. These trace identities serve as
useful checks on the accuracy of the computed spectrum of $M$: identities involving
higher powers of $M$ can also be derived, involving plaquette (or even higher loop)
sums, but will not be needed here. Of course, ${\rm det}(M)={\rm det}(D)$.

  For not too large lattices, the Lanczos algorithm can be used to extract the
complete spectrum of the hermitian matrix $M$: for a review of the issues involved
in computing complete spectra with Lanczos, we refer the reader to \cite{loop}.
In this paper we shall consider 5-dimensional lattices dimensioned $L^{4}$x$L_{5}$
with $L=$4,6 and $L_{5}$=12. For the case $L=$4, the matrix $M$ has dimension 12$VL_{5}$=36864,
and resolving the complete spectrum to 9 significant figures requires on the
order of 200,000 Lanczos sweeps. One then finds that ${\rm Tr}(M)\simeq$10$^{-9}$ and
the left and right hand sides of (4) agree to 9 significant figures. The computational
effort required grows as the square of the dimension of $M$, however, and on larger
lattices one eventually encounters missed eigenvalues (due to accidental close
degeneracies) which make this brute-force approach to extracting a full spectrum, and
determinant, ultimately impractical. However, having an ``exact" result for the
domain-wall determinant (at least, to 9 significant figures) will turn out to be
a useful check on the accuracy of the stochastic methods described below.

\section{Stochastic Procedure for Determinant Ratios}

   The stochastic formalism developed by Golub and coworkers \cite{golub} provides
an approach to the calculation of determinant ratios which allows us to go to
larger, and more physically realistic lattices. This technique was first applied to
the quark determinant problem in lattice QCD by Irving and Sexton \cite{Sexton}.
In this approach, given a positive
definite hermitian matrix $P$, the close connection between Lanczos recursion and
Gaussian integration is exploited to compute an arbitrary diagonal matrix element
$<v|f(P)|v>$ of any differentiable function $f(P)$ of the matrix $P$. Any such 
matrix element can be written as a spectral sum which can then be interpreted as a
Riemann-Stieltjes integral computable via the usual variety of Gaussian quadrature
rules (Gauss, Gauss-Radau, Gauss-Lobatto, etc.). It turns out that the appropriate
weights for the Gaussian quadrature are directly related to the spectrum of the
tridiagonal matrix generated by Lanczos recursion on $P$ (for further details, see
\cite{golub}). With a technique for computing diagonal matrix elements in hand,
this approach then leads immediately to a stochastic method for unbiased estimates
of the ${\rm Tr}(f(P))$ as the average over an ensemble of diagonal expectations
of $f(P)$ with respect to a set of random vectors $|z_{i}>$ where each vector
has components $\pm 1$ chosen at random:
\begin{eqnarray}
   {\rm Tr}(f(P)) &=& E(<z|f(P)|z>)  \\
    var(<z|f(P)|z>) &=&  2\sum_{i\neq j} |f(P)_{ij}|^{2}
\end{eqnarray}

  The above method can be fruitfully applied to our problem- the evaluation of 
$\frac{{\rm det}(M(U,A))}{{\rm det}(M(U,0))}$- as follows. Choosing
\begin{equation}
   P \equiv M(U,0)^{-1}M(U,A)^{2}M(U,0)^{-1}
\end{equation}
we then have a positive definite hermitian matrix $P$ as required for the algorithm
described above (for the work described in this paper, we have performed
the required inversions of $M(U,0)$ using the minimum residual algorithm
\cite{numrec2}). Choosing $f(P) = \ln{(P)}$, we obtain the (log of the square of the) desired
determinant ratio.
Moreover, for sufficiently weak electromagnetic fields $A$
it is clear that $P$ is close to the identity, which has the following salutary 
consequences:
\begin{enumerate}
\item As we shall see shortly, the Lanczos recursion to compute the individual 
diagonal matrix elements $<z| \ln{(P)} |z>$ converges {\em extremely rapidly}:
in fact 4 or 5 Lanczos sweeps (corresponding to 8 or 10 inversions of 
the domain wall matrix $M$) are typically sufficient to obtain the matrix
element to single precision accuracy. 
\item As $P$ is close to the identity, (6) implies that the variance of the 
diagonal matrix elements should also be small, allowing us to obtain a good
estimate of the trace with an ensemble of manageable size.
\end{enumerate}

  We have studied these issues on a variety of small lattices, generating 
pure SU(3) quenched configurations (at $\beta$=6.0) as the strong interaction
background field, and Coulomb gauge photon field configurations corresponding
to electric charge $e_{q}=-0.1$ (roughly, the down quark charge). On a 4$^4$x12
lattice we then typically find 
$\ln{\frac{{\rm det}(M(U,A))}{{\rm det}(M(U,0))}}= \frac{1}{2} {\rm Tr}\ln{(P)}\simeq -5$.
For a specific diagonal element $<z|\ln{(P)}|z>$ with a typical  random vector $|z>$, the
convergence of the Lanczos estimates is indicated in Table 1, both for 4$^4$x12
and for 6$^4$x12 lattices. In either case, the convergence to seven or
eight significant figures occurs after less than 10 Lanczos sweeps, with the rapidity
of convergence basically insensitive to the lattice volume. The computational
effort required for the Lanczos recursion is of course dominated by the inversions
of the domain wall operator $M(U,0)$.

\begin{table}[htb]
\centering
\renewcommand{\baselinestretch}{1.0}
\caption{Convergence of Lanczos estimates for a diagonal element of $\ln{(P)}$}
\vspace{.1in}
\label{tab:convla}
\begin{tabular}{||c|c|c||} \hline
 Lanczos sweep &  $<z|\ln{(P)}|z>,L=4$ & $<z|\ln{(P)}|z>,L=6$\\
\hline
2 &   -11.29817352 &  -53.44524316  \\
3 &   -11.54470058 &  -55.10595906  \\
4 &   -11.54923739 &  -55.14478821  \\
5 &   -11.54931955 &  -55.14572594  \\
6 &   -11.54932110 &  -55.14575078  \\
7 &   -11.54932113 &  -55.14575146  \\
8 &   -11.54932113 &  -55.14575148  \\ 
\hline
\end{tabular}
\end{table}
\renewcommand{\baselinestretch}{1.7}

\begin{figure}
\psfig{figure=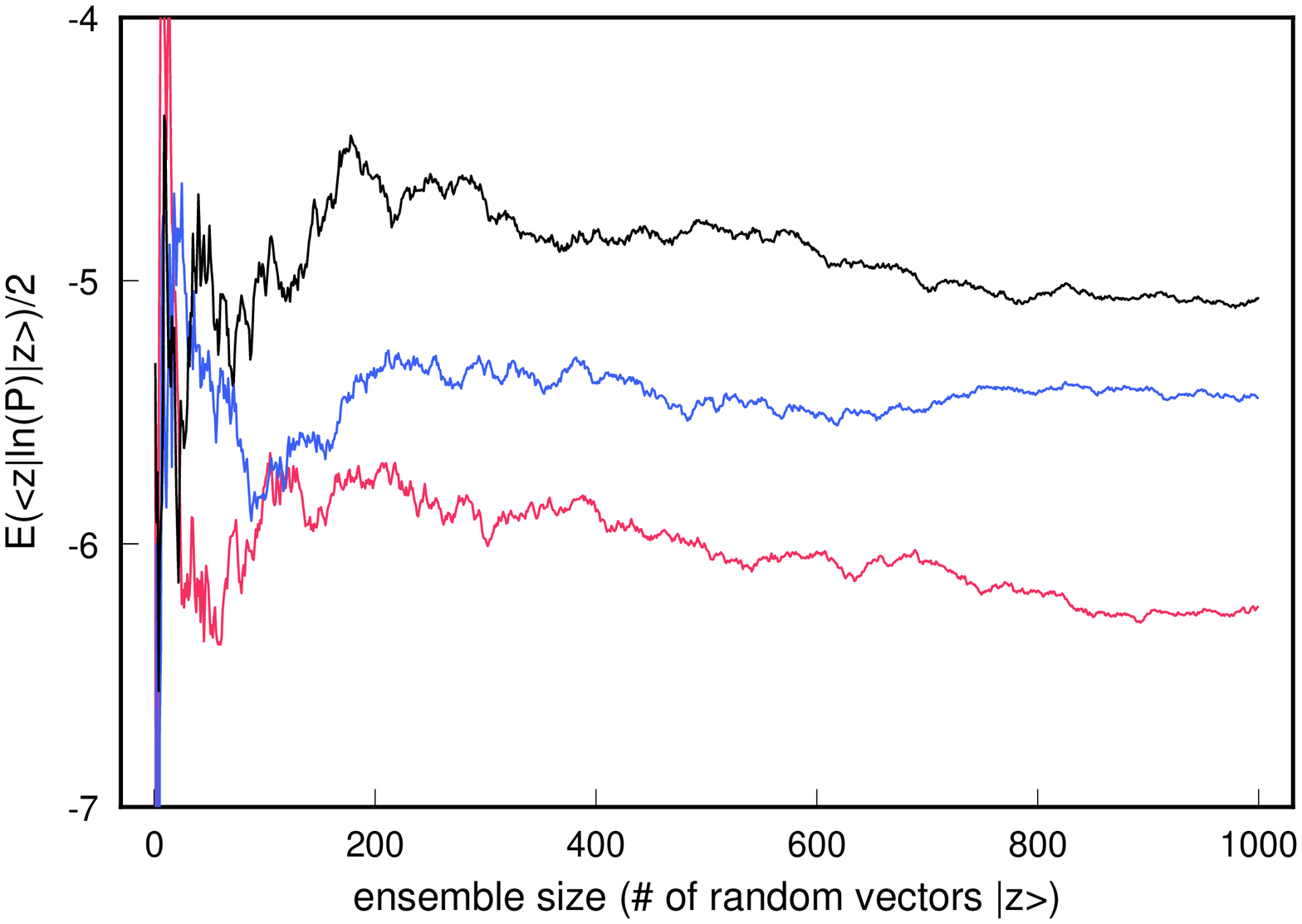,height=0.5\hsize}
\caption{Ensemble convergence for log(determinant ratio),varying photon field, 4$^4$x12}
\end{figure}   

 The computation of the determinant ratio, or equivalently the trace of $\ln{P}$, requires
repeating the evaluation of the diagonal matrix element in a random vector $|z>$ many
times, but this part of the procedure is of course trivially parallelizable as the
random vectors (and resulting estimates of the trace) are completely independent. We show in
Figure 1 the rate of convergence of the desired determinant ratio $E(<z|\ln{(P)}|z>)/2$
as a function of the ensemble size, up to a final sample size of 1000 random vectors, for
three statistically independent photon field $A$ superimposed on the same background QCD
field $U$. For the red curve, the exact (log)ratio of determinants, evaluated by computation
of the complete spectrum as described in the preceding section, gives -6.1522, while the
stochastic result (again, for 1000 random vectors) is -6.2384$\pm$0.1414. For the blue
curve, the comparison is -5.3791 to -5.4472$\pm$0.1346, and for the black curve -5.0609 to
-5.0635$\pm$0.1388. On a 4$^4$x12 lattice, each evaluation of $<z|f(P)|z>$ in (5)
takes about 20 seconds on a Xeon 2.8GHz PC.
 Note that the variation of the log determinant ratio is of order unity
between different configurations. In Figure 2 the same convergence issues are displayed,
this time varying both the abelian $A$ and nonabelian $U$ background fields.

\begin{figure}
\psfig{figure=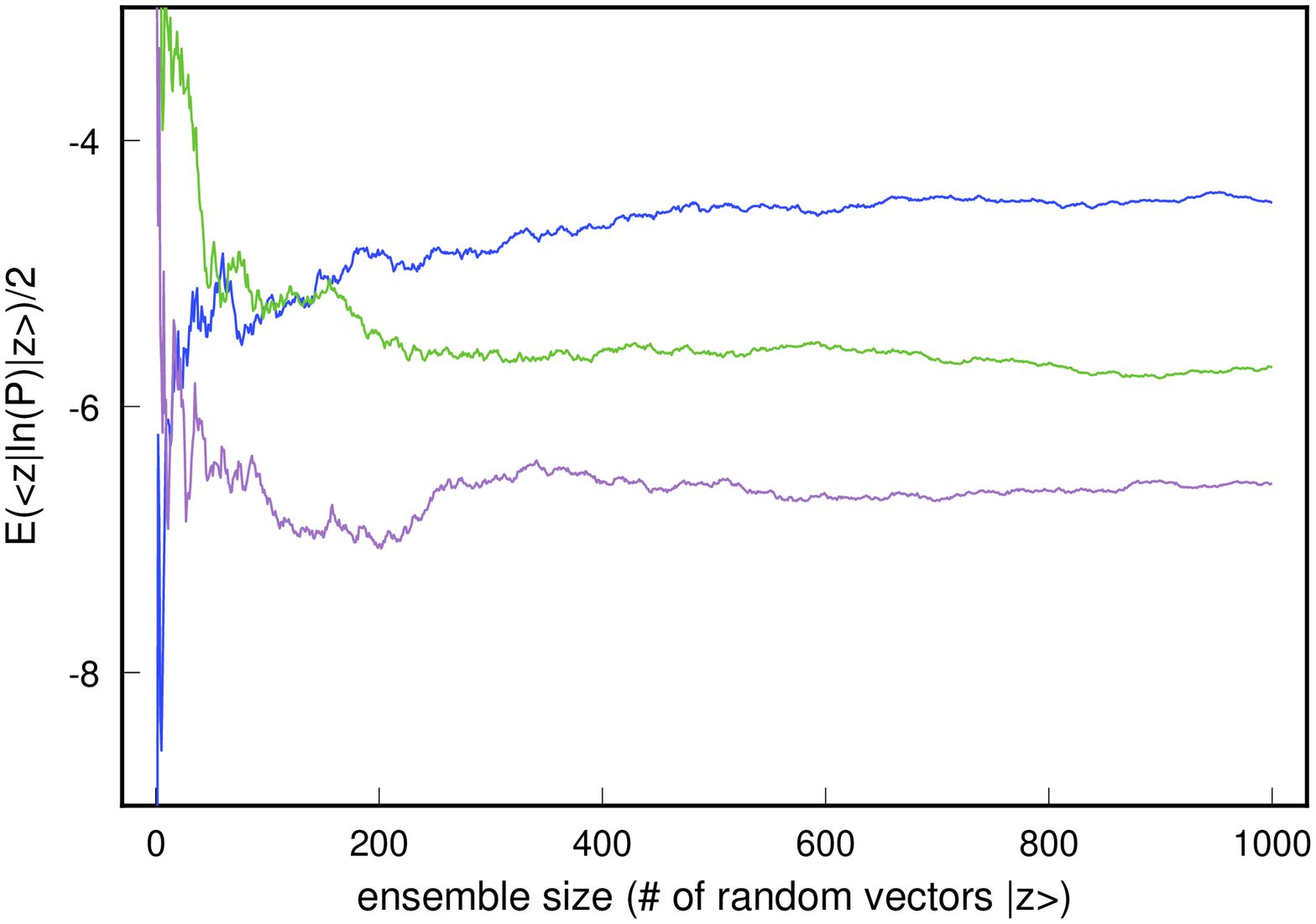,height=0.5\hsize}
\caption{Ensemble convergence for log(determinant ratio),varying photon,gluon fields, 4$^4$x12}
\end{figure}   

  We have also repeated the stochastic determination of determinant ratios for
a variety of 6$^4$x12 lattices (with the same mass and coupling parameters as 
above), and Figure 3 shows the convergence from an ensemble of 4000 estimates
for three different U(1) fields superimposed on a fixed QCD background. The
variation of the log(determinant ratio) from one U(1) field to another again is
of order unity. The
total variance (6) turns out to be linear in the volume of the lattice: for
$L$=4 we find a variance per site of 0.0063(2), while for $L$=6 the variance
per site is 0.0066(5). The computed log(determinant ratio) is also of course
an extensive quantity proportional to the lattice volume, so we conclude that
we can achieve a fixed relative error in the log(determinant ratio) evaluation
with a fixed ensemble size, or a fixed absolute error by increasing the
number of stochastic estimates linearly with the lattice volume. Thus the
absolute error on the log(determinant ratio) 
achieved for the 6$^4$x12 lattices with 4000 estimates is about 0.16, comparable
to the accuracy achieved with 1000 estimates on the smaller lattices (which are
about one fifth the volume).
As emphasized
previously, the evaluations of $<z|f(P)|z>$ in (5) for different random
vectors $|z>$ are completely independent so this average is trivially
parallelizable with 100\% efficiency.

\begin{figure}
\psfig{figure=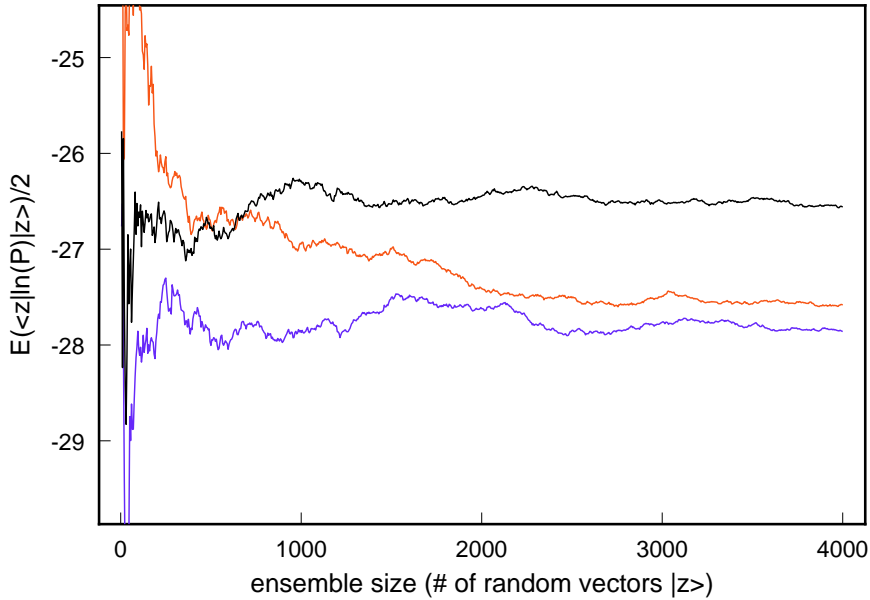,height=0.5\hsize}
\caption{Ensemble convergence for log(determinant ratio),varying photon field, 6$^4$x12}
\end{figure}

\section{Computational Issues}

  We have described two different routes to computing the electromagnetic
variation of the quark determinant in unquenched QCD with domain wall
quarks: the direct, ``brute force" evaluation of the complete spectrum
of the (hermitian) domain wall operator via Lanczos recursion, or a
stochastic, approximate evaluation using the Golub-Bai-Fahey method.
The first approach, in which the Lanczos recursion is carried out in
double precision, delivers a result for the determinant (provided all
eigenvalues can be resolved) which is typically accurate to a precision
5 or 6 significant places short of double precision, with the loss of
accuracy due to the need for distinguishing spurious from real eigenvalues
\cite{Cull}. The computational effort scales superficially as $V^2$
for lattice volume $V$ (where the first factor of volume arises from
the multiplication by $M(U,A)$ at each Lanczos sweep, and the second from the
fact that evaluation of the complete spectrum requires a number of Lanczos
sweeps greater than and roughly proportional to $V$). However, for larger lattices
the eigenvalue density rises to the point where double precision Lanczos
recursions will be unable to resolve many of the eigenvalues, and the method
simply becomes impractical in the absence of hardware-implemented extended
precision arithmetic. Moreover, the most efficient techniques for diagonalization
of the tridiagonal Lanczos matrix (e.g. the QL algorithm with implicit shifts
\cite{NumRec}) are intrinsically serial by nature, which makes parallelization
of this part of the procedure difficult (although less efficient parallel
algorithms, based on the Sturm sequence property \cite{GolubBook}, have been used
successfully for small lattices). The stochastic approach described in the
previous section also scales as $V^2$ with the lattice volume, if we require
a fixed absolute error in the log(determinant ratio), but does not require 
extended precision arithmetic as we go to larger lattices.  Moreover, 
parallelization of this approach is trivial, as pointed out above. For the lattices
studied here ($L$=4,6,$L_5$=12) the two methods are comparable in computational
effort (if we require an error in the stochastic approach at the few percent level),
so we conclude that the stochastic approach is probably the appropriate technique
for larger systems. 

\newpage
\section{Acknowledgements}
The work of A. Duncan  was supported in part by 
NSF grant 010487-NSF. A.D. is grateful for the hospitality of the
Max Planck Institut f\"ur Physik (Werner Heisenberg Institut), where
part of this work was done. The work of E. Eichten was performed at
the Fermi national Accelerator Laboratory, which is operated
by University Research Association, Inc., under contract DE-AC02-76CH03000.

\end{document}